# Creativity and the New Structure of Science

Andrei P. KIRILYUK[*]

Institute of Metal Physics, Kiev, Ukraine 03142

A qualitatively new, much more liberal and efficient organisation of science is proposed and justified in connection with emerging international science structures, such as the European Research Council, and growing debates about further role and development of fundamental science.

Increasing international cooperation in science today constitutes an important part of the growing globalisation tendency and leads to intense activity of science organisation at the emerging superior level of explicitly globalised knowledge. A fresh example is provided by the proposition to create the European Research Council (ERC) [1,2] that would monitor the fundamental science development in Europe, complementing various existing institutions and tools, which are oriented mainly to practical, technological applications of science [3]. However, fundamental science organisation and purposes form a subject of vivid discussion in recent years revealing a number of intrinsic, deeply rooted problems [4-39], and it would be quite relevant to take into account the emerging conclusions and to choose only those solutions for the new, international science organisation that would not reproduce (and inevitably amplify) its well-known limitations at the previous (basically national) level. In this connection one should not forget an essential difference between the applied and fundamental science development: while the former is naturally guided by the powerful pressure of immediate practical purposes, the latter is largely left with its own, internal "criteria of truth" and estimates (referring only to hypothetical, less predictable consequences in a remote enough future), which multiply increases the danger of various spurious, subjective influences in fundamental science, leading to impasses and harmful practical effects, especially in the current situation of socially high stakes and globally amplified ambitions.

It is a well-known fact (see e.g. [3-6,12,13,24,27,38,39]) that practical organisation of fundamental science is dominated now by the rigidly fixed and highly centralised (pyramidal) kind of structure, or *unitary system*, which reproduces the main properties of the administrative, or command, type of economic organisation within a totalitarian system of power (exemplified by the recently fallen Communist regime in the former Soviet Union and its ideological satellites).[1] The "central power" decides eventually everything in that kind of organisation (though most often *not* by *direct* orders), contrary to mainly "personal" (independent) kind of elementary decisions and the "distributed", *dynamically emerging* evaluation of their results in the non-totalitarian, liberal (or "free-market") system, whose dynamics is always more chaotic *in the details*. The unitary formal structure of science is efficiently supported also by less formal, emerging tools of centralised control, such as the notorious "peer review" system [10-16] that determines eventually every real action (publication, financial support) in fundamental science and is considered for the moment as a key instrument also for the new national or international institutions, such as ERC [1,2]. It does not need to be proved that practical efficiency of the unitary system is intrinsically low and cannot be increased without changing its nature for a liberal type of structure. Whereas the economic version of this conclusion has been progressively recognised almost everywhere, in the fundamental science structure we remain strangely at the level of the most centralised regimes of the past and continue to suffer unnecessarily from the inevitable destructive consequences of that outdated "intellectual totalitarianism" exerted in various, explicit and implicit, forms [4-6,10-16,21-24,27-30,34-39].[2]

---

[*] Address for correspondence: Post Box 115, Kiev-30, Ukraine 01030. E-mail address: kiril@metfiz.freenet.kiev.ua.

[1] Those who still yield to the "officially" maintained illusion of "liberal" dynamics of modern science organisation can have a closer look at its practical details, where a large enough *Central Committee*, realised as a top-level bureaucratic monster (a National Academy, Science Foundation, a Continental Ministry, etc.), serves to impose to "masses" of totally *dependent* "ordinary" researchers practical decisions that directly involve the essential science *content* and are actually produced by a much more narrow group of top priests, *Politburo*, including a formal "Board", but even more a few of informally related authorities, remaining hidden behind the formal structures, but actually promoting arbitrary decisions in their personal favour. Often presenting itself as a kind of *meritocracy* ("rule of the best"), the self-chosen scientific mafia quickly degenerates into pretentious, but vain *mediocracy* and then to spoiled merdocracy, with the only result of omnipresent destruction of the very essence of science. While making reference, similar to all totalitarian regimes, to intense "consultation with the people", the privileged top priesthood looks exclusively for prolongation of its parasitic dominance and increase of unmerited profits by reproduction of the same kind of unitary subordination at ever smaller scales within science structure.

But similar to social system case, the time comes when the current unitary system, because of its intrinsic deficiency, cannot maintain and develop its activity any more, even if the system is not subjectively challenged from the outside. Today we have attained just that "critical state" in European (and world) science, irrespective of material prosperity or technological successes. The sacramental "to be or not to be?" quickly loses its rhetorical flavour and acquires the rough practical meaning of dramatic degradation and effective disappearance of the whole large and once flourishing disciplines, especially among fundamental and "exact" sciences, such as various branches of physics.[3] The long stagnation in their essential development (absence of fundamental "discoveries") leads inevitably to considerable loss of public interest (especially on the background of *purely empirical*, technological successes), which in turn provokes further decrease of creativity, thus closing the self-amplifying "positive feedback loop" of destructive tendency that quickly grows and invades the whole system. Numerous and detailed discussions of this and other aspects of modern situation in fundamental science can easily be found in a variety of serious enough sources [5-37] which cannot be suspected of artificially organised subjective attack, especially if one takes into account the accompanying much larger body of informally transmitted opinions and attitudes. But whereas particular criticism of the existing science organisation grows "exponentially", the root of the problem is not clearly specified in a unified manner and therefore the proposed solutions are reduced either to "treatment of symptoms", i. e. illusive "amelioration" of the evidently outdated, fundamentally irrelevant kind of science organisation, or to mechanistic support of its occurring degradation presented as a new, progressive "mode" (cf. e. g. [24-26]), despite the evident drop of quality.

---

[2] It is not difficult to see that the same conclusions refer, in principle, to the whole structure of state power, or more generally "public institutions", including most "democratic", *formally* liberal versions of their organisation. Indeed, today that traditional organisation of "common services" becomes also inefficient and dangerously degrading in all its results, which only confirms the serious necessity to consider other possible kinds of organisation that could be conveniently tested just for the case of science organisation, remaining relatively independent of other state functions. In any case, the unitary structure in science is especially outdated and harmful with respect to other public services because *efficient* science is an *intrinsically creative*, individually structured and unpredictable in detail, kind of activity, the property that contradicts directly and permanently the unitary general organisation of state. A good scientist is always bad in the role of a standard state "clerk" or "official" (and still worse as a "bureaucrat"), and the reverse. Whereas the "traditional", unitary organisation of science is inherited from previous epochs of hot and cold wars, where one of the main arguments in favour of science at a "national" level has been creation of ever more powerful weapons and other "strategically" important tasks (national energy sources, communication means, etc.), the emerging more and more "globalised", inter-dependent continent and world structure is incompatible with fundamental scientists being considered as small soldiers of big armies engaged in a military fight with each other (certain current attempts to provoke an artificial return to the "old good times" of the "hot", and bloody, history only demonstrate the deep crisis of the conventional, unitary organisation of society and civilisation).

[3] A visibly much more prosperous state of "soft", especially biological sciences should not be misleading: these branches of fundamental knowledge only repeat, in a compressed form, the previous evolution of physics, starting from an impressive series of "epochal" discoveries and ending in successful applications that leave no place for creative development of fundamental knowledge. The latter is certainly necessary as persisting "mysteries" and growing inconsistencies in physics convincingly demonstrate, but it is possible only within a qualitatively new kind of science organisation at a "high-tech" civilisation stage. The conventional, unitary science organisation will inevitably tend to strongly limit its development to purely empirical, short-sighted technology support: the hardest barrier for development is always created by a successful development itself, and the unitary organisation of science is intrinsically unable to propose a way out of that auto-impasse, while the conceptually blind, but empirically omnipotent technology necessarily becomes fatally destructive (cf. [40]).

A more radical, but well substantiated and quality-guided cure I would like to defend here can be understood by analogy to the transition from the command to liberal economy, but is provided also with a deep scientific basis in terms of the new, universal concept of complexity, where it is described as the emerging, inevitable ascent to a "higher complexity level" of system dynamics by a "generalised phase transition" [38]. This rigorous, scientific interpretation only confirms the evident, common-sense application of the above economic analogy and states that the *exhausted* unitary, centralised type of organisation should and can be *progressively* replaced now by a *qualitatively* different, much more variable, highly *interactive* community of (usually) small and *independent* enterprises, tentatively designated here as Creative Research Units (CRU), each of them constituting the *minimum* self-sufficient unit of professional scientific research *or management* related to science. The proper research units are engaged in scientific problem solution as such, whereas its evaluation and presentation for various funding opportunities is performed by separate enterprises of management which obtain their part from successful project funding (or, say, from unmerited funding they reveal by their specially attributed function) and *compete* among them for larger numbers of more successful results. Total professional and public *openness* of each project results, their evaluation and use is another feature of the new structure of science, closely related to the independent, liberal operation mode of the main structural units (similar to the *regulated* market economy). A variety of CRUs with different, and suitably changing, *dynamically adaptable* purposes and sizes will *naturally*, "spontaneously" emerge and evolve, at a national, European, and world scale, similar to the dynamics of developed, sustainable market structure. This transition to the *superior complexity* of science structure is *inevitable* because of the increasingly and now critically growing complexity in the science *content* and the universal law of *correspondence, or symmetry, of complexity* [38,39] rigorously confirming the practically obvious requirement of correspondence between complexity of the system *form* (organisation) and *content* (it is now *critically* missing in the unitary science structure).

In particular, initiation of several enterprises of the new kind and the corresponding "interactive" monitoring within the future ERC activity represents a convenient opportunity for introduction of the new kind of science structure in Europe and elsewhere. Certain "germs" of the new kind of structure spontaneously appear (and grow), by the way, already within the unitary system (and often in opposition to it), in the form of various, more or less independent, science "foundations" and special, usually private, "institutes" (cf. e. g. [13,24]), but they cannot fully realise the potentialities of the new kind of organisation they actually represent because of the absence of other essential members of the community (such as management enterprises) and their provisional, unstable status with respect to the dominating unitary system. On the other hand, various informal "groups of support" of particular initiatives ("lobbies") emerge within the unitary system management, but they also preserve their formally "underground" status and restricted efficiency of final results (where today's difficulties in advancement of any new idea about the content or structure of science provide many real examples of those limitations). In all those situations one deals with "subcritical seeds" of the new structure, in terms of the physics of structure formation, in which an intuitively transparent result is rigorously demonstrated, namely that the appearing small germs of the new, objectively favoured structure should nevertheless possess a big enough (though generally small) size in order to ensure its further stable growth. One can say thus that creation and support of a number of small, but "above-critical" seeds of the new structure of science, as it is outlined here, can be seen as an important practical purpose of science development within new initiatives, such as ERC, which can provide an additional, *conceptually* strong support and meaning for those initiatives themselves.

The new system is intrinsically oriented towards *explicit creation* in research [38,39], which is appreciated most and practically determines everything else, contrary to the domination of formal "positions" in the unitary system, inevitably brought to a collapse because of the inherent contradiction between its structure and formally announced "high" purposes. In particular, due to the knowledge market dynamics itself, the efficiency of new knowledge production is naturally and *independently* estimated by *qualitative* properties of the results, i. e. the *explicit, consistent*, and thus *certainly* useful *problem solutions*, as opposed to apparently senseless, but inevitably dominating quantitative estimates of the "intensity of attempts" in the self-seeking unitary structure. The creative scientist is now liberated from the routine "paper work" imposed by the exhausting fight for "positions": instead, he is actually "attacked" by requests to do this work for him, coming from competing management enterprises which look anxiously for "promising" projects, the main source of their existence (but the same reasons will also force them to reject, or considerably improve, low-quality projects that constitute a potential source of their losses). The *intrinsic creativity* of genuine science finds thus its new and *unlimited* realisation within the new kind of organisation, in close relation to the intrinsically high, *sustainable quality* of results, whereas the unitary system is inevitably opposed to any true novelty and hence creative, high-quality science, the latter remaining, however, the unique source of subjective "interest" of both scientists and "general public" (giving, in particular, each next generation of scientists). The resulting living, self-developing knowledge market will naturally eliminate the dangerously growing separation of "professional" science from other spheres of activity, "practical life", "amateur scientists", and the related "public interest", which is another "irresolvable", inevitably growing problem within the unitary science structure [17-26,35-39].

Note that the conventional, commercial market, though being liberal intrinsically, operates within a unitary kind of social structure (including the most "democratic" versions of unitary system), whose modern realisation is closely related to the *industrial*, massive and simplifying, way of production and life style. This omnipresent "industrialisation" is objectively opposed to the unreduced creativity of the formally liberal market dynamics and largely kills its "evolutionary" advantages, especially at the modern stage of the "developed" (democratic) unitary system (cf. e.g. ref. [41]). In this respect, the proposed non-unitary science organisation is designed to avoid not only authoritarian centralisation of power, but also its "liberal" version appearing as industrial, standardising and complexity-destroying approach, in science as well as in material production. Such qualitative change of tendency is possible due to the intrinsically creative character of scientific activity, now liberated from the formal subordination to the unitary bureaucratic hierarchy of imposed "governors". In that way one performs also the necessary transition from the industrial, severely reduced and eventually destructive "liberty", to the unreduced creativity of interactive CRUs, provided the underlying basic principles outlined above are maintained in practical initiation of the emerging new system (that becomes then autonomously, *dynamically sustainable* above a critical size/age).

And finally, it is important that both rigorous justification of transition to the qualitatively new, nonunitary kind of science structure and the experience of practical transition realisation are naturally and directly extendible to other spheres of "common-service" activity, with the evident relevance to the currently emerging "critical" situation in unitary governance of *any* public and private structures of all scales (cf. popular discussions about "open society" [24,41] or "non-governmental organisations" and the related problems of development [42]). The role of Europe and its intellectual elites in this larger, inevitably forthcoming transition to the new structure of society [38] must be active and "decision-making", if the current tendency of stagnation and decline is ever to be replaced by the intrinsic creation.

**Acknowledgement**

The author gratefully acknowledges the support by Euroscience of his participation in the recent conference "New Science and Technology Based Professions in Europe" (Bischenberg, France, 6-9 November 2002, http://www.euroscience.org/WGROUPS/YSC/bischenberg2002.htm) which stimulated the appearance of this article (although its ideas remain at the author's responsibility and do not seem to be divided by many). Discussion of the article content with Françoise Praderie (Observatoire de Paris) has contributed to the presented result formulation.

# References

[1] "New research council needed", *Nature* **418** (2002) 259.

Q. Schiermeier, "A window of opportunity", *Nature* **419** (2002) 108.

I.B. Holland, "Europe is not yet ready for a research council", *Nature* **419** (2002) 248.

W. Krull, "A fresh start for European science", *Nature* **419** (2002) 249.

"New structures for the support of high-quality research in Europe", ESF position paper, April 2003, http://www.esf.org/newsrelease/63/ERC.pdf.

[2] J.P. Connerade, "Will Europe have its research area?", *Euroscience News*, No. 22, Winter (2003) 1. Electronic version is accessible at http://www.euroscience.org/MBSHIP/Bulletin/Euro22.pdf.

F. Sgard, "The view of Euroscience", *Euroscience News*, No. 22, Winter (2003) 4.

W. Krull, "Climbing the steps towards a European Research Council", *Euroscience News*, No. 22, Winter (2003) 5.

G. Schatz, "Do we need a European Research Council? A scientist's view", *Euroscience News*, No. 22, Winter (2003) 6.

O. Postel-Vinay, "European Research Council expected for 2004", *TRN*, July 1 (2003), http://trn.rampazzo.com/W22Y03/leader.asp.

[3] See e. g. the official European Commission's research website including the rules for the forthcoming Sixth Framework Programme, http://europa.eu.int/comm/research/fp6/index_en.html. For a discussion, see also: "Frameworks can be too rigid", *Nature* **420** (2002) 107.

[4] T. Kuhn, *The Structure of Scientific Revolutions* (Chicago University Press, 1970). First edition: 1962.

[5] L. de Broglie, "Nécessité de la liberté dans la recherche scientifique", In *Certitudes et Incertitudes de la Science* (Albin Michel, Paris, 1966). Original edition: 1962.

L. de Broglie, "Les idées qui me guident dans mes recherches", In *Certitudes et Incertitudes de la Science* (Albin Michel, Paris, 1966). Original edition: 1965.

[6] G. Lochak, *Louis de Broglie. Un prince de la science* (Flammarion, Paris, 1992).

[7] J. Maddox, "Restoring good manners in research", *Nature* **376** (1995) 113.

J. Maddox, "The prevalent distrust of science", *Nature* **378** (1995) 435.

[8] J. Ziman, "Is science losing its objectivity?", *Nature* **382** (1996) 751.

[9] J. Horgan, *The End of Science. Facing the Limits of Knowledge in the Twilight of the Scientific Age* (Addison-Wesley, Helix, 1996).

J. Horgan, "From Complexity to Perplexity", *Scientific American*, June (1995) 74.

J. Horgan, *The Undiscovered Mind: How the Human Brain Defies Replication, Medication, and Explanation* (Touchstone/Simon & Schuster, New York, 1999).

[10] A. Sangalli, "They burn heretics, don't they?", *New Scientist*, 6 April (1996) 47.

[11] Y. Farge, "Pour davantage d'éthique dans le monde de la recherche", *Science Tribune*, October 1996, http://www.tribunes.com/tribune/art96/farg.htm.

[12] A.A. Berezin, "Hampering the progress of science by peer review and by the "selective" funding system", *Science Tribune*, December 1996, http://www.tribunes.com/tribune/art96/bere.htm.

[13] D. Braben, "The repressive regime of peer-review bureaucracy?", *Physics World*, November (1996) 13.

[14] C. Wennerås and A. Wold, "Nepotism and sexism in peer-review", *Nature* **387** (1997) 341.

[15] K. Svozil, "Censorship and the peer review system", e-print physics/0208046 at http://arXiv.org.

[16] D. Adam and J. Knight, "Publish, and be damned ...", *Nature* **419** (2002) 772.

[17] J. Brockman, *The Third Culture: Beyond the Scientific Revolution* (Touchstone Books, 1996). Online edition can be found at http://www.edge.org/documents/ThirdCulture/a-TC.Cover.html.

J. Brockman, "The Third Culture", *Edge*, http://www.edge.org/3rd_culture/index.html.

*The New Humanists: Scientists at the Edge*, edited by J. Brockman (Barnes & Noble Books, 2003). See also *Edge*, http://www.edge.org/3rd_culture/brockman/brockman_print.html.

[18] *The Flight from Science and Reason*, eds. P.R. Gross, N. Levitt, and M.W. Lewis (New York Academy of Sciences, 1996). Also: *Annals of the New York Academy of Sciences*, vol. **775**.

[19] J. Bricmont, "Le relativisme alimente le courant irrationnel", *La Recherche*, No. 298, mai (1997) 82.

[20] A. Sokal and J. Bricmont, *Impostures intellectuelles* (Odile Jacob, Paris, 1997).

[21] J. de Rosnay, "Du pasteur au passeur", *Le Monde de l'Education, de la Culture et de la Formation*, No. 245, fevrier (1997) 20.

[22] O. Postel-Vinay, "La recherche menacée d'asphyxie", *Le Monde de l'Education, de la Culture et de la Formation*, No. 245, fevrier (1997) 47.

O. Postel-Vinay, "La défaite de la science française", *La Recherche*, No. 352, avril (2002) 60.

O. Postel-Vinay, "L'avenir de la science française", *La Recherche*, No. 353, mai (2002) 66.

[23] *Sciences à l'école: De désamour en désaffection*, *Le Monde de l'Education*, Octobre (2002), 25-42.

[24] S. Fuller, *The Governance of Science: Ideology and the Future of the Open Society* (Open Society Press, Buckingham, 1999).

S. Fuller, *Thomas Kuhn: A Philosophical History for Our Times* (University of Chicago Press, 2000).

[25] M. Gibbons, "Science's new social contract with society", *Nature* **402** Supp (1999) C81.

[26] H. Nowotny, P. Scott, and M. Gibbons, *Rethinking Science: Knowledge and the Public in an Age of Uncertainty* (Polity Press, 2001).

M. Gibbons, H. Nowotny, C. Limoges, M. Trow, S. Schwartzman, and P. Scott, *Evolution of Knowledge Production: The Dynamics of Science and Research in Contemporary Societies* (Sage, London, 1994).

[27] D.S. Greenberg, *Science, Money, and Politics: Political Triumph and Ethical Erosion* (University of Chicago Press, 2001).

[28] M. Lopez-Corredoira, "What is research?", e-print physics/0201012 at http://arXiv.org. *Ciencia Digital* 8 (2000).

M. Lopez-Corredoira, "What do astrophysics and the world's oldest profession have in common?", e-print astro-ph/0310368 at http://arXiv.org.

[29] C. Chiesa and L. Pacifico, "Patronage lies at the heart of Italy's academic problems", *Nature* **414** (2001) 581.

[30] P.A. Lawrence, "Rank injustice", *Nature* **415** (2002) 835.

P.A. Lawrence, "The politics of publication", *Nature* **422** (2003) 259.

"Challenging the tyranny of impact factors", *Nature* **423** (2003) 479.

P.A. Lawrence and M. Locke, "A man for our season", *Nature* **386** (1997) 757.

[31] G. Brumfiel, "Misconduct in physics: Time to wise up?", *Nature* **418** (2002) 120.

"Reputations at risk", Editorial, *Physics World*, August (2002).

E. Check, "Sitting in judgement", *Nature* **419** (2002) 332.

[32] L. Stenflo, "Intelligent plagiarists are the most dangerous", *Nature* **427** (2004) 777.

[33] M.V. Simkin and V.P. Roychowdhury, "Copied citations create renowned papers?", e-print cond-mat/0305150 at http://arXiv.org.

M.V. Simkin and V.P. Roychowdhury, "Stochastic modeling of citation slips", e-print cond-mat/0401529 at http://arXiv.org

[34] M. Gad-el-Hak, "Publish or Perish – An Ailing Enterprise?", *Physics Today* **57**, March (2004) 61. Accessible at http://physicstoday.org/vol-57/iss-3/p61.html.

[35] S. Nagel, "Physics in Crisis", *FermiNews* **25**, No. 14, August 30 (2002), http://www.fnal.gov/pub/ferminews/ferminews02-08-30/p1.html.
Also: *Physics Today* **55**, September (2000) 55.

P. Rodgers, "Hanging together", *Physics World*, October (2002), Editorial.

[36] K.H. Elliott, "Physics with Everything", Editorial, *Eur. J. Phys.* **24**, 3 March (2003).

[37] S. Sengupta, "Science as a Culture – Its Implications", e-print physics/0310161 at http://arXiv.org.

[38] A.P. Kirilyuk, *Universal Concept of Complexity by the Dynamic Redundance Paradigm: Causal Randomness, Complete Wave Mechanics, and the Ultimate Unification of Knowledge* (Naukova Dumka, Kiev, 1997), 550 p., in English. For a non-technical review see also: e-print physics/9806002 at http://arXiv.org.

[39] A.P. Kirilyuk, "Dynamically Multivalued, Not Unitary or Stochastic, Operation of Real Quantum, Classical and Hybrid Micro-Machines", e-print physics/0211071 at http://arXiv.org. See especially section 9 for a detailed discussion of modern science structure and development.

[40] M. Rees, *Our Final Hour: A scientist's warning: How terror, error, and environmental disaster threaten humankind's future in this century* (Basic Books, New York, 2003).

[41] G. Soros, *Open Society: Reforming Global Capitalism* (Public Affairs Press, New York, 2000).

G. Soros and G. Shandler, *The Crisis of Global Capitalism: Open Society Endangered* (Public Affairs Press, New York, 1998).

G. Soros, "Toward a Global Open Society", *The Atlantic Monthly* **281**, January (1998) 20. Accessible at http://www.theatlantic.com/issues/98jan/opensoc.htm.

[42] *Terms for Endearment: Business NGOs and Sustainable Development*, ed. by J. Bendell (Greenleaf Publishing, 2000).